\documentclass[10pt,final,a4paper,twocolumn]{IEEEtran}
\usepackage{verbatim} 
\usepackage{graphicx}
\usepackage{epstopdf}
\usepackage{multirow}
\usepackage{multicol}
\usepackage{amsmath}
\usepackage{latexsym}
\usepackage{amssymb}
\usepackage{tcolorbox}
\usepackage{enumerate}
\usepackage{cuted}
\usepackage[switch]{lineno}

\newcommand{\bd}{\mathbf}
\newcommand{\dg}{^{\mathrm o}}
\newcommand{\uI}{^{\mathrm I}}
\newcommand{\uC}{^{\mathrm C}}
\newcommand{\us}{^{\mathrm s}}
\newcommand{\cosa}{c_{\alpha}}
\newcommand{\cose}{c_{\epsilon}}
\newcommand{\cosp}{c_{\rho}}
\newcommand{\sina}{s_{\alpha}}
\newcommand{\sine}{s_{\epsilon}}
\newcommand{\sinp}{s_{\rho}}
\newcommand{\lx}{_{\mathrm x}}
\newcommand{\ly}{_{\mathrm y}}

\IEEEoverridecommandlockouts
\begin{document}
\title{Camera Calibration Using Inaccurate and Asynchronous Discrete GPS Trajectory from Drones}
\author{R. Yang\\
Y. Bar-Shalom\\
H.A.J. Huang\\
\today\\

\thanks{Manuscript received November 14, 2022; revised January 24, 2023; re-
leased for publication May 29, 2023.}
\thanks{Refereeing of this contribution was handled by Florian Meyer.}
\thanks{Y. Bar-Shalom, Department of ECE, University of Connecticut, Storrs, CT 06269, USA (E-mail: yaakov.bar-shalom@uconn.edu); R. Yang and H.A.J. Huang, DSO National Laboratories, 12~Science Park Drive, Singapore~118225 (E-mails: yrong@dso.org.sg, hhongan@dso.org.sg).}
}
\markboth{JAIF}%
{Shell \MakeLowercase{\textit{et al.}}: Bare Demo of IEEEtran.cls for Journals}

\maketitle

\begin{abstract}
This paper considers a stationary camera calibration problem, which estimates the camera orientation angles yaw, pitch and roll, using a drone trajectory recorded by a GPS. There are three challenges in using a GPS trajectory as ground truth for camera calibration. One, the altitude of GPS data is inaccurate with an unknown bias. Two, the GPS receiver and camera are not time synchronized, and there is an unknown time offset between the two systems. Three, the GPS trajectory is time-discrete and accurate interpolation is needed. This is actually an estimation problem since velocity is also needed. To address the first two challenges, we formulate the problem as a parameter estimation problem to estimate a vector consisting of the GPS altitude bias and time offset in addition to the camera yaw, pitch and roll biases. We then develop a special maximum likelihood  estimator using the Iterated Least Squares  algorithm which can work with a non-synchronized time-discrete GPS trajectory for the third challenge. Since the camera measurement errors are usually small, this requires a high calibration accuracy so that the residual bias error following the calibration should not be significant compared to the measurement error standard deviation. The calibration accuracy depends highly on the drone trajectory. This paper also recommends an appropriate drone trajectory which can yield a good calibration accuracy, namely, 14\% of the measurement error standard deviation. Simulation tests are conducted to demonstrate the algorithm performance. The estimation results meet the Cramer-Rao Lower Bound (CRLB) since the Normalized Estimation Error Squared  w.r.t.\ the CRLB is statistically acceptable.
\end{abstract}
\section{Introduction}
\label{s1}
This paper presents a camera calibration approach for a stationary camera which looks at air targets. We assume the camera is of ``pinhole" type without radial and tangential distortion. The position of the camera is assumed known. The calibration computes the camera orientation, which is defined by three rotation angles, yaw, pitch and roll. Since the camera is looking for air targets, there is no fixed object with known position in its Field of View (FOV). The calibration is based on the trajectory of a drone instrumented with a GPS receiver, which, however, usually has a significant altitude bias error\footnote{The altitude estimate from GPS is substantially less accurate than the horizontal position since there are fewer high-orbit satellites, which provide most of the altitude information, vs.\ low orbit satellites, which provide most of the horizontal location information.}. Also the camera and GPS receiver are not time synchronized. This introduces an unknown fixed time offset between the GPS and camera time-stamps. Furthermore, the drone trajectory is a sequence of discrete points with a certain time interval and there is no analytical expression for the trajectory. This paper will develop a practical approach for the problem of estimating the camera orientation, the GPS altitude bias and the time offset.

Camera calibration is not a new problem. Numerous works have been done before, and they can be categorized into two areas: computer vision related applications and estimation theory based approaches. The camera calibration in computer vision is developed from the Perspective-n-Point (PnP) problem \cite{Fischler1981}\cite{Kneip2011}. The original PnP problem is described as follows: Given the relative spatial locations of $n$ control points $P_i$ with $i=1,\ldots,n$, and given the angle to every pair of these points from an additional point, called the center of perspective $C$, find the lengths of the line segments joining $C$ to each of the control points. The camera calibration is based on the matching of $n$ 3D control points and their corresponding points on the 2D image space. They share the same angles of arrival with reference to the camera center of perspective $C$. A number of solutions have been developed with this approach \cite{Fischler1981}\cite{Linnainmaa1988}\cite{Haralick1991}\cite{Haralick1994}\cite{Kneip2011}. Some focused on the solution of the minimum number of control points required ($n=3$) as P3P problem~\cite{Linnainmaa1988}\cite{Haralick1991}\cite{Haralick1994}\cite{Kneip2011}, and some deal with a large number of points consisting of outliers and inaccurate points. The RANSAC~\cite{Fischler1981} scheme can be applied to select good samples. Some extensions on the camera calibration take unknown focal length and radial distortion into consideration~\cite{{Josephson2009}}\cite{Wu2015}. 

If we apply the PnP approach to our problem, a 3D GPS-instrumented drone trajectory needs to match the  camera-measured 2D trajectory. Since there is an unknown time offset between GPS-based 3D and camera 2D trajectories and an unknown altitude bias on the 3D trajectory, it is not practical to apply point-to-point 3D-2D matching. We then seek a solution from the estimation theory.

Unlike the computer vision approach, which develops the camera calibration as a particular geometric problem, the estimation theory approach formulates the camera calibration as a parameter estimation problem with stochastic models. It defines the unknown parameter to be estimated as $\theta$, and builds a relationship between $\theta$ and measurements that include noise. If the problem is observable (i.e., with a unique solution), optimization algorithms, such as Gradient Descent, Newton's algorithm or Iterated Least Squares (ILS)~\cite{B7}, can be applied in a systematic manner. A number of works along this line have been carried out to estimate the sensor position/orientation and measurement biases. This is often referred to as the sensor registration problem. It can be solved offline using either ILS or Maximum Likelihood (ML) estimator from a batch of data~\cite{458}\cite{Zhou1997}\cite{Zhou1999}\cite{Okello2003}\cite{Fortunati2011}\cite{Zhang2000}, or online (estimating the sensor biases and target trajectories simultaneously) using a Kalman Filter (KF) type dynamic estimator or Recursive Least Squares (RLS) approach~\cite{BarShalom11}\cite{Lin2004}\cite{Yang2010}\cite{Wilthil2016}. The online approaches (also referred as to auto-calibration) sound more attractive. However, they estimate a large augmented state consisting of all target states and sensor biases. This large state may create computational infeasibility for real time when the number of targets is large. Furthermore, sensor bias estimation accuracy is not always guaranteed as arbitrary target trajectories do not reduce the bias error compared to a dedicated special trajectory. The calibration accuracy (or sensor bias estimation accuracy) is paramount in our problem as camera orientation must be accurately estimated so that the residual bias error should not be significant compared to the camera measurement error. We therefore prefer an offline approach which allows a GPS-equipped drone to fly in a special predefined path dedicated to accurate camera calibration. Such a path will be discussed in the sequel.
The previous work on offline sensor registration mainly dealt with radar pose and measurement biases~\cite{Zhou1997}\cite{Zhou1999}\cite{Okello2003}\cite{Fortunati2011}. 
Camera calibration was conducted in~\cite{458}\cite{Zhang2000}. The yaw, pitch and roll biases of multiple cameras and target locations are estimated simultaneously using the ILS method in \cite{458} for a satellite-based camera observing an exoatmospheric target of opportunity. In \cite{Zhang2000}, a camera was calibrated through observing a planar pattern shown at several different orientations, and camera intrinsic and extrinsic parameters were estimated using a closed-form solution. Neither of them deals with unknown time offset among different systems, for example, sensor and ground truth systems  --- the different sensors are assumed time synchronized. 

Online and offline calibration with an unknown time offset have been discussed in various applications~\cite{Li2004}\cite{Kelly2010}\cite{Mair2011}\cite{Furgale2013}\cite{Li2014}\cite{Li2020}\cite{Rehder2016}. We focus on the offline solutions~\cite{Kelly2010}\cite{Mair2011}\cite{Furgale2013}\cite{Rehder2016}. In~\cite{Kelly2010}\cite{Mair2011} the time offset and spatial calibration were conducted separately in sequence. The time offset was estimated first, and then spatial calibration was conducted. A more robust approach~\cite{Furgale2013}\cite{Rehder2016} estimated the time offset and spatial biases simultaneously. This was a robotics application with camera, Inertial Measurement Unit (IMU) and laser rangefinder. It estimated the time offset among sensors and measurement transformation. However, the camera was assumed well calibrated. The Levenberg-Marquardt (LM) algorithm was used to minimize an objective function based on the maximum likelihood criterion, using stationary objects detected by the camera and rangefinder on a moving platform. Although the approach included unknown time offsets into its estimation parameter, camera calibration was not conducted.

In this paper, we develop our approach based on estimation theory which will include the GPS altitude bias and time offset in the estimation of the parameter vector $\theta$. Another challenge is that the GPS 3D trajectory is given in numerical form. The preliminary version of the present study, \cite{Yang2022}, conducted calibration assuming an accurate GPS without altitude bias. An ILS algorithm was developed to perform calibration based on a stochastic model dealing with a GPS trajectory expressed by a sequence of discrete-time points. In the present paper, inaccurate GPS with unknown altitude bias is used. The calibration accuracy drops significantly with this additional unknown unless it is part of the estimated parameter vector. If the residual bias error (following the calibration) is not small enough compared to the camera measurement error, the calibration is not meaningful. We will develop an enhanced ILS algorithm to improve the estimation accuracy, and recommend a practical drone path to achieve good calibration accuracy.  

The rest of paper is structured as follows. Section~\ref{s2} describes the three coordinate systems used in this paper. Section~\ref{s3} describes the problem formulation, namely, the stochastic model for estimation. Section~\ref{s4} presents the estimation algorithm based on the stochastic model dealing with numeric GPS trajectories. Section~\ref{s5} presents simulation results on calibration error, and recommends a suitable drone path. Section~\ref{s6} draws the conclusions.

\section{Coordinate systems}
\label{s2}
\begin{figure}[ht]
	\begin{center}\leavevmode
		\includegraphics[width=3.4in]{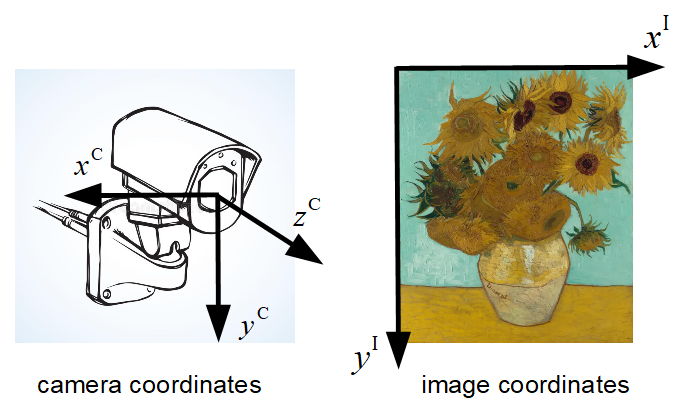}
	\end{center}
	\caption{Camera and image coordinate systems.}
	\label{f1}
\end{figure}

The following three coordinate systems are used in this paper:
\begin{itemize}
    \item Common coordinate system with $x$-$y$-$z$ as East, North and Up (ENU).
    \item Camera coordinate system with $x\uC$-$y\uC$-$z\uC$ centered at the camera position $(x\us,y\us,z\us)$, shown in Fig.~\ref{f1}.
    \item Image coordinate system with $x\uI$-$y\uI$ shown in Fig.~\ref{f1}.
\end{itemize}

The notations used in the paper are listed in Table~\ref{tb_1}.
\begin{table}[!h]
	\caption{Notations}
	\label{tb_1}
	\centering
	\begin{tabular}{l|l}
	\hline \hline
    ${\bd x}$ & $[x\ y\ z]'$, a point in the common (ENU) coordinate system.\\
    ${\bd x}\uC$ & $[x\uC\ y\uC\ z\uC]'$, a point in the camera coordinate system.\\
    ${\bd x}\uI$ & $[x\uI\ y\uI]'$, a point in the image coordinate system.\\
    ${\bd x}\us$ & $[x\us\ y\us\ z\us]'$ sensor (camera) position.\\
    $\alpha$ & camera pointing azimuth or yaw (clockwise from N).\\
    $\epsilon$ & camera pointing elevation or pitch (up from horizontal).\\
    $\rho$ & camera roll (ideally zero), clockwise around the center of\\ & the Focal Plane Array (FPA).\\
    $\hbar$ & GPS altitude bias. The GPS provided altitude is higher than \\& the true value when $\hbar$ is positive, otherwise, $\hbar$ is negative.\\
    $\tau$ & time offset between the drone GPS and the camera. The GPS\\
    &clock is ahead of the camera clock when $\tau$ is positive,\\
    &otherwise, $\tau$ is negative.\\\hline
	\end{tabular}
\end{table}

The conversion of $\bd x$ to $\bd x\uC$ is given by  
\begin{eqnarray}
    \bd x\uC &=& \bd T(\alpha,\epsilon,\rho) (\bd x-\bd x\us)\nonumber\\
    &=& \bd T^z(\rho) \bd T^x(\epsilon-90\dg) \bd T^z(-\alpha)(\bd x-\bd x\us)
\label{e2_1}
\end{eqnarray}
where we use the following mnemonic notations for rotations between 3D Cartesian systems:

\begin{equation}
\bd T^x(\phi) = \left[\begin{array}{ccc} 1&0&0 \\
0 & \cos \phi&  \sin \phi \\ 
 0 & -\sin \phi &  \cos \phi \end{array}\right]
 \label{e2_2}
\end{equation}
for a rotation around the $x$-axis by $\phi$ from $y$ toward $z$,
\begin{equation}
\bd T^z(\phi) = \left[\begin{array}{ccc} \cos \phi & \sin \phi & 0 \\ 
-\sin \phi & \cos \phi & 0 \\ 
0&0&1 \end{array}\right]
 \label{e2_3}
 \end{equation}
for a rotation around the $z$-axis by $\phi$ from $x$ toward $y$. The rotation around the $y$-axis is not necessary as $\bd T^x(90\dg-\epsilon)$ replaces the $y$-axis by the $z$-axis, so that rotation around the $z$-axis occurs twice. The combined rotation in (\ref{e2_1}) is
\begin{equation}
    \bd T(\alpha,\epsilon,\rho) = \left[\begin{array}{ccc}
        \cosa\cosp+\sina\sine\sinp & \cosa\sine\sinp-\sina\cosp & -\cose\sinp\\
        \sina\sine\cosp-\cosa\sinp & \sina\sinp+\cosa\sine\cosp & -\cose\cosp\\
        \sina\cose & \cosa\cose & \sine
    \end{array}\right]
\label{s2d}
\end{equation}
where 
\begin{align}
    &\sina=\sin \alpha, \hspace{0.1in} \sine=\sin \epsilon, \hspace{0.1in} \sinp=\sin \rho\\
    &\cosa=\cos \alpha, \hspace{0.1in} \cose=\cos \epsilon, \hspace{0.1in} \cosp=\cos \rho
\end{align}

The conversion of $\bd x\uC$ to $\bd x\uI$ is
\begin{equation}
    \bd x\uI = \bd f(\bd x\uC) = \left[\begin{array}{l}
     \frac{P\lx}{2}+\frac{x\uC f}{z\uC} \\ \\
      \frac{P\ly}{2}+\frac{y\uC f}{z\uC}
\end{array} \right]
\label{e2_4}\end{equation}
where $f$ is the focal length with units of pixel (assumed square)
\begin{equation}\label{e2_5}
    f=\frac{P\lx}{2\tan(\Theta\lx/2)}=
    \frac{P\ly}{2\tan(\Theta\ly/2)}
\end{equation}
and $P\lx$ and $P\ly$ are the numbers of pixels in $x\uI$ and $y\uI$ coordinates, respectively; $\Theta\lx$ and $\Theta\ly$ are the fields of view (FOV) --- angular spans --- in $x\uI$ and $y\uI$, respectively.

\section{Problem Formulation}
\label{s3}

This section formulates the estimation problem in a stochastic model. The parameter to estimate is
\begin{equation}
    \mathbf{\theta}=[\alpha\ \epsilon\ \rho\ \hbar\ \tau]'
\label{e3_1}
\end{equation}
which consists of three camera orientation angles $\alpha$, $\epsilon$ and $\rho$, GPS altitude bias~$\hbar$, and the time offset between the drone GPS and camera systems~$\tau$, which are estimated simultaneously. The stochastic model for estimating $\mathbf{\theta}$ is
\begin{equation}\label{e3_2}
    \bd Z=\bd H(\bd \theta,\bd X)+\bd w
\end{equation}
where $\bd H(\cdot)$ is defined in (\ref{e3_7}), $\bd Z$ is the camera measurement vector consisting of $n$ discrete-time points in the image coordinates as
\begin{eqnarray}\label{e3_3}
     \bd Z&=&[\bd z(t_1)'\ \ldots\  \bd z(t_n)']'\nonumber\\
     &=&[\bd x\uI(t_1)'\ \ldots\  \bd x\uI(t_n)']'
     +\bd w
\end{eqnarray}
with measurement times $t_1, \ldots, t_n$, $\bd w$ is a $2n$ zero-mean Gaussian measurement noise vector with covariance 
\begin{equation}
\bd R = \bd I_{2n \times 2n} \sigma_{\rm F}^2
\label{e3_4}\end{equation}
and $\sigma_{\rm F}^2$ is the variance of the measurement noise in the FPA. For details of how this is obtained based on the optics' Point Spread Function (PSF) and pixel size, see~\cite{543}. $\bd X$ is the GPS drone trajectory (with unknown altitude bias and time offset) represented by a set of discrete-time points in the common coordinate system (ENU) at times corresponding to the camera measurement times, corrected by the (unknown) time offset. It is defined as
\begin{figure}[ht]
	\begin{center}\leavevmode
		\includegraphics[width=3.5in]{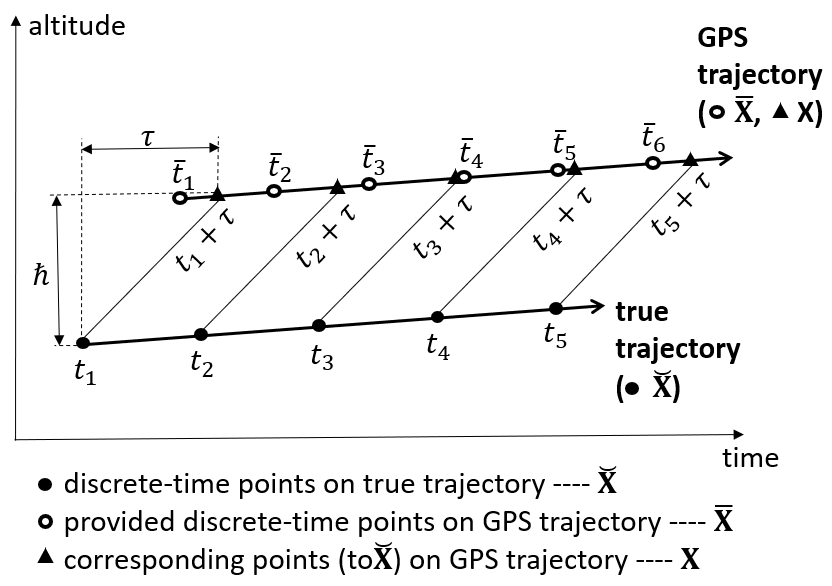}
	\end{center}
	\caption{True and GPS trajectories.}
	\label{f2}
\end{figure}
\begin{equation}
    \bd X=[\bd x(t_1+\tau)'\ \ldots\  \bd x(t_n+\tau)']'
\label{e3_5}
\end{equation}
However, $\bd X$ is not known exactly. The available information on the GPS trajectory is
\begin{equation}
    \overline{\bd X}=[\bd x(\overline t_1)'\ \ldots\  \bd x(\overline t_m)']'
\label{e3_6}
\end{equation}
where $\overline t_1, \ldots, \overline t_m$ do not correspond to the times in $\bd X$, and $\overline{\bd X}$ and $\bd X$ intervals can differ. We need to find the relationship between $\overline{\bd X}$ and $\bd X$, so that the model in~(\ref{e3_2}) can be utilized for estimation. This will be solved in next section. 

Fig.~\ref{f2} shows the relationship of the true trajectory $\breve{\bd X}$ and GPS trajectory of the drone, where
\begin{equation}
    \breve{\bd X}=[\breve{\bd x}(t_1)'\ \ldots\  \breve{\bd x}(t_n)']'
\end{equation}
Each discrete-time point ($\bullet$) on the
true trajectory has a corresponding point ($\blacktriangle$) on the GPS
trajectory. The relationship of the $i$th points of $\breve{\bd X}$ and $\bd X$ is
\begin{equation}
    \breve{\bd x}(t_i)=\bd x(t_i+\tau)-[0\ 0\ \hbar]'
\end{equation}
The measurement function $\bd H$ in~(\ref{e3_2}) is then
\begin{eqnarray}
    \bd H&=&\left[\begin{array}{c}
         \bd h_1[\alpha,\epsilon,\rho,\breve{\bd x}(t_1)]\\
         \vdots\\
         \bd h_n[\alpha,\epsilon,\rho,\breve{\bd x}(t_n)]
    \end{array}\right]
\label{e3_7}
\end{eqnarray}
 with
\begin{align}  
    \bd h_i(\cdot)&=\bd f\left\{\bd T(\alpha,\epsilon,\rho)[\overline{\bd x}(t_i+\tau)-[0\ 0\ \hbar]'-\bd x\us]\right\}\nonumber\\
    &=\bd f(\bd x\uC_i)=\bd x\uI_i\nonumber\\
&\hspace{1.6in} i=1,\ldots,n \label{e3_8}
\end{align}
The above converts a position ``$\blacktriangle$" $\bd x_i$ to a position ``$\bullet$" $\breve{\bd x}_i$, then converts to camera coordinates as $\bd x\uC_i$ using~(\ref{e2_1}), and finally converts to the image space as $\bd x\uI_i$ using~(\ref{e2_4}).

\section{Estimation Algorithm}
\label{s4}
\begin{figure}[ht]
	\begin{center}\leavevmode
		\includegraphics[width=2.5in]{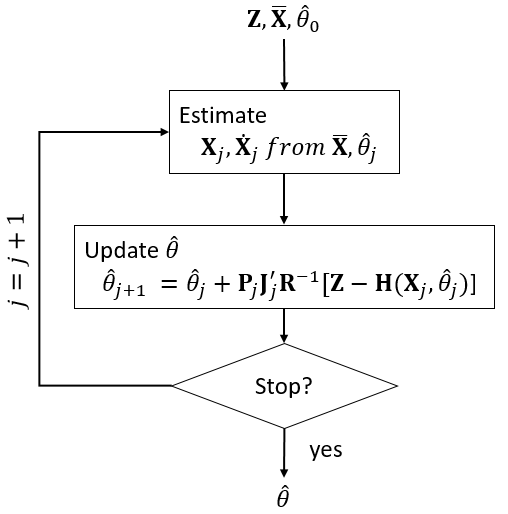}
	\end{center}
	\caption{The ILS estimation algorithm.}
	\label{f2a}
\end{figure}

This section solves the problem described in Section~\ref{s3} using a unique Iterated Least Squares (ILS) algorithm which is illustrated in Fig.~\ref{f2a}. 
Its uniqueness lies in the fact that $\bd Z$ and $\overline{\bd X}$ are used to estimate $\bd X$ and
$\dot{\bd X}$ and then $\bd Z$ and $\bd X$ are used in the iterative estimation of $\mathbf{\theta}$, defined in (\ref{e3_1}).

Given the camera measurement $\bd Z$, the GPS trajectory $\overline{\bd X}$ and the initial value of the parameter $\hat{\theta}_0$, the algorithm finds $\hat{\theta}$ through iteration, indexed by $j$, based on the nonlinear model given in~(\ref{e3_2}). We will describe the algorithm with the following three steps:
\begin{enumerate}[(A)]
    \item Estimation of $\bd X_j$ and its velocity~$\dot{\bd X}_j$ from~$\overline{\bd X}$ and~$\hat{\theta}_j$ in the $j$th iteration. $\bd X_j$ and $\dot{\bd X}_j$ are needed in the $\theta$ estimation in (B);
    \item Updating of $\hat{\theta}_j$ to $\hat{\theta}_{j+1}$ using an optimization algorithm based on the model~(\ref{e3_2}) with estimated $\bd X_j$ and~$\dot{\bd X}_j$;
    \item Stop the iteration when a satisfactory $\hat{\theta}$ is obtained.
\end{enumerate}

\subsection{Estimate $\bd X$ and its Velocities $\dot{\bd X}$}
\label{s41}
\begin{figure}[ht]
	\begin{center}\leavevmode
		\includegraphics[width=3.0in]{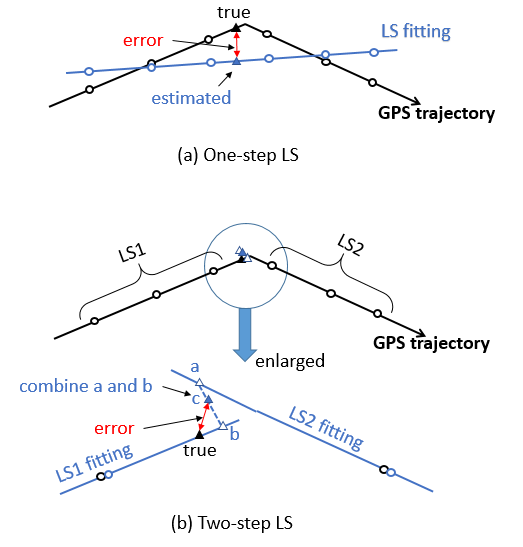}
	\end{center}
	\caption{The LS fitting algorithms. (a) One-step LS fitting approach developed in~\cite{Yang2022}. (b) Two-step LS fitting approach used in the present paper.}
	\label{f2b}
\end{figure}

To estimate $\theta$ we need to know $\bd X$ and its velocities $\dot{\bd X}$ from the positions $\overline{\bd X}$ (namely, to estimate the GPS trajectory ``$\blacktriangle$" points from ``$\circ$" points in Fig.~\ref{f2}), so that the discrete-time points on the GPS trajectory are at times $[t_1+\tau, \ldots, t_n+\tau]$ corresponding to the camera measurements at times $[t_1, \ldots, t_n]$.\footnote{This amounts to more than interpolation since the velocities are also estimated and a special approach is used when the drone maneuvers.} The Least Squares (LS) fitting algorithm developed in~\cite{Yang2022} used a sliding window containing the neighbouring ``$\circ$" points before and after a particular ``$\blacktriangle$" to estimate its position and velocity. 
However, this will not perform well when a maneuver happens within the window. We therefore enhance it as a two-step LS fitting approach in this paper. Fig.~\ref{f2b} illustrates the two steps. We can see the one-step LS approach in (a) has a large error when there is a maneuver. The two-step LS fitting approach shown in (b) uses two LS estimators on the neighbouring points before and after the ``$\blacktriangle$". They obtain two estimates ``b" and ``a", respectively. The final estimate ``c" is a combination of ``b" and ``a". The estimation error of the two-step LS fitting is therefore reduced significantly.

In the two-step LS fitting approach, we illustrate LS1 (applied to the neighbors before the ``$\blacktriangle$") to obtain point ``b" in detail in the following. The estimation of point ``a" is similar. First of all,  we estimate the velocities and accelerations of the nearest ``$\circ$" [assuming $\bd x(\overline t_i)$] before the  ``$\blacktriangle$". Its velocity and acceleration are
    \begin{eqnarray}
        \dot{\bd x}(\overline t_i) &=& [\dot x(\overline t_i)\ \dot y(\overline t_i)\ \dot z(\overline t_i)]'\label{e4_30}\\
        \ddot{\bd x}(\overline t_i) &=& [\ddot x(\overline t_i)\ \ddot y(\overline t_i)\ \ddot z(\overline t_i)]'\label{e3_31}
    \end{eqnarray}
The vectors consisting of velocities and accelerations in x,y and z coordinates are defined as
    \begin{eqnarray}  
        \bd d_x^i&=&[\dot x(\overline t_i)\ \ddot x(\overline t_i)]'\\
        \bd d_y^i&=&[\dot y(\overline t_i)\ \ddot y(\overline t_i)]'\\
        \bd d_z^i&=&[\dot z(\overline t_i)\ \ddot z(\overline t_i)]'
    \end{eqnarray}
They are estimated separately. The model to estimate $\bd d_x^i$ from its neighbors is
    \begin{equation}
        \bd \Delta_x^i=\bd D^i \bd d_x^i\\
    \label{e46}
    \end{equation} 
where
    \begin{eqnarray}
        \bd \Delta_x^i&=&\left[\begin{array}{c} x(\overline t_{i-\delta})-x(\overline t_i) \\
        \vdots\\
        x(\overline t_{i-1})-x(\overline t_i)
        \end{array}\right]\\
        \bd D^i&=&
        \left[\begin{array}{cc}
        \overline T_{i-\delta} & -0.5\overline T^2_{i-\delta}\\
        \vdots & \vdots\\ 
        \overline T_{i-1} & -0.5\overline T^2_{i-1}
        \end{array}\right]
    \end{eqnarray}
and 
    \begin{eqnarray}
        \overline T_{i-k}&=&\overline t_{i-k}-\overline t_i\\
        &&\hspace{0.3in} k=[\delta,\ldots,1]\nonumber
    \end{eqnarray}
The number of neighboring points used in~(\ref{e46}) is~$\delta$=3. LS is applied to estimate $\bd d_x^i$ as
    \begin{equation}
        \hat{\bd d}_x^i=[(\bd D^i)' \bd D^i]^{-1}(\bd D^i)'\bd \Delta_x^i
    \end{equation}
and $\bd d_y^i$ and $\bd d_z^i$ are estimated similarly as
     \begin{eqnarray}
        \hat{\bd d}_y^i&=&[(\bd D^i)' \bd D^i]^{-1}(\bd D^i)'\bd \Delta_y^i\\
        \hat{\bd d}_z^i&=&[(\bd D^i)' \bd D^i]^{-1}(\bd D^i)'\bd \Delta_z^i        
    \end{eqnarray}
Next, we compute positions and velocities of ``$\blacktriangle$", namely, ``b" point in Fig.~\ref{f2b}(b). We assume the ``$\blacktriangle$" is the $k$th point in $\bd X$. The positions and velocities are computed by
    \begin{align}
        \left[\begin{array}{c}
           x_b(t_k+\tau_j) \\
            \dot{x}_b(t_k+\tau_j)
        \end{array}
        \right]
        =&
        \left[\begin{array}{ccc}
            1 & T_k & \dfrac{T_k^2}{2}\\
            0 & 1 & T_k
        \end{array}
        \right]
        \left[\begin{array}{c}
        x(\overline t_i)\\
        \dot{x}(\overline t_i)\\
        \ddot{x}(\overline t_i)
        \end{array}\right]\\
        \left[\begin{array}{c}
            y_b(t_k+\tau_j) \\
            \dot{y}_b(t_k+\tau_j)
        \end{array}
        \right]
        =&
        \left[\begin{array}{ccc}
            1 & T_k & \dfrac{T_k^2}{2}\\
            0 & 1 & T_k
        \end{array}
        \right]
        \left[\begin{array}{c}
        y(\overline t_i)\\
        \dot{y}(\overline t_i)\\
        \ddot{y}(\overline t_i)
        \end{array}\right]\\
        \left[\begin{array}{c}
            z_b(t_k+\tau_j) \\
            \dot{z}_b(t_k+\tau_j)
        \end{array}
        \right]        =&
        \left[\begin{array}{ccc}
            1 & T_k & \dfrac{T_k^2}{2}\\
            0 & 1 & T_k
        \end{array}
        \right]
        \left[\begin{array}{c}
        z(\overline t_i)\\
        \dot{z}(\overline t_i)\\
        \ddot{z}(\overline t_i)
        \end{array}\right]
    \end{align}
where $T_k=t_k+\tau_j-\overline t_i$, and $\tau_j$ is from  $\theta_j$ in the $j$th iteration. The likelihood of point “b” [see in Fig. 4(b)] is computed using the measurement residual
\begin{equation}
     \bd v_b=[(\bd \Delta_x^i-\bd D^i \bd d_x^i)'\ \ (\bd \Delta_y^i-\bd D^i \bd d_y^i)'\ \ (\bd \Delta_z^i-\bd D^i \bd d_z^i)']' 
\end{equation}
according to
\begin{equation}
    \mathcal L_b=\mathcal N(\bd v_b;\bd 0,\bd I)
\end{equation}
where $\mathcal N(\cdot)$ is the standard $3\delta$-multivariate Gaussian pdf.

The second step LS is computed in a similar manner to obtain the positions, velocities and likelihood of point ``a". The final estimate, for point ``c"  in Fig.~\ref{f2b}(b) is based on a weighted average as follows:
    \begin{align}
        \left[\begin{array}{c}
            \hat x(t_k+\tau_j)\\
            \hat{\dot{x}}(t_k+\tau_j)\\
            \hat y(t_k+\tau_j)\\
            \hat{\dot{y}}(t_k+\tau_j)\\
            \hat z(t_k+\tau_j)\\
            \hat{\dot{z}}(t_k+\tau_j)\\
        \end{array}
        \right]
        =& \dfrac{\mathcal L_a}{\mathcal L_a+\mathcal L_b}
        \left[\begin{array}{c}
            x_a(t_k+\tau_j)\\
            \dot{x}_a(t_k+\tau_j)\\
            y_a(t_k+\tau_j)\\
            \dot{y}_a(t_k+\tau_j)\\
            z_a(t_k+\tau_j)\\
            \dot{z}_a(t_k+\tau_j)\\
        \end{array}
        \right]\nonumber\\
        &+ \dfrac{\mathcal L_b}{\mathcal L_a+\mathcal L_b}
        \left[\begin{array}{c}
            x_b(t_k+\tau_j)\\
            \dot{x}_b(t_k+\tau_j)\\
            y_b(t_k+\tau_j)\\
            \dot{y}_b(t_k+\tau_j)\\
            z_b(t_k+\tau_j)\\
            \dot{z}_b(t_k+\tau_j)\\
        \end{array}
        \right]
    \end{align}

\subsection{Update the estimate of $\theta$}
\label{s42}
The parameter given in~(\ref{e3_1}) is estimated based on
\begin{equation}
\hat{\mathbf{\theta}}= \arg\min_{\mathbf{\theta}}
||\bd Z - \bd H(\bd \theta,\bd X)||^2_{{\bd R}^{-1}}
\label{e4_1}\end{equation}
Using the ILS \cite{B7} to solve the above optimization\footnote{The ILS is the numerical algorithm to solve for the ML Estimate under Gaussian assumption.}, one has
\begin{eqnarray}
    \hat{\bd \theta}_{j+1}&=&\hat{\bd{\theta}}_j+\bd P_j\bd J_j'\bd R^{-1}[\bd Z-\bd H(\bd \theta_j,\bd X_j)]\label{e4_2}\\
    \bd P_j&=&(\bd J_j'\bd R^{-1}\bd J_j)^{-1}\label{e4_3}\\
    &&\hspace{1.5in} j=1,\ldots,n_j\nonumber
\end{eqnarray}
with the Jacobian
\begin{equation}
    \bd J_j=[\nabla_{\bd \theta_j} \bd H(\bd \theta_j,\bd X)']'
    =[\nabla_{\bd \theta_j} \bd h_1(\cdot)'\ \ldots\ \nabla_{\bd \theta_j} \bd h_n(\cdot)']'
\label{e4_4}\end{equation}
where $j$ is the iteration index. The final estimate $\hat{\mathbf{\theta}}$ is the value to which the iteration~(\ref{e4_2}) converged using a stopping criterion. The derivatives needed for~(\ref{e4_4}) are given in Appendix~\ref{a2}. 

\subsection{Stopping Criterion}

To obtain a good calibration result, we set a tight stopping criterion. First, we normalize the measurement residual squared element by element in iteration~$j$
\begin{equation}
    \bd V_j = [\bd Z - \bd H(\bd X_j,\hat{\theta}_j)]\otimes[\bd Z - \bd H(\bd X_j,\hat{\theta}_j)]  \sigma_{\rm F}^{-2}
\end{equation}
Then, we check every element $v_{j,i}$ with $(i=1\ldots 2n)$ in $\bd V_j$, where $2n$ is the number of measurements times the measurement dimension 2. All $v_{j,i}$ must be below the ``3 sigma" limit
\begin{equation}
    v_{j,i} \leq 3^2
\end{equation}
This element-wise checking criterion can prevent a few large measurement residuals being smoothed by a large number of small residuals. Also, to prevent a run that cannot meet the stopping criterion, the maximum number of iterations is set to~20.

\section{Simulation Results}
\label{s5}
\begin{figure}[ht]
	\begin{center}\leavevmode
		\includegraphics[width=3.4in]{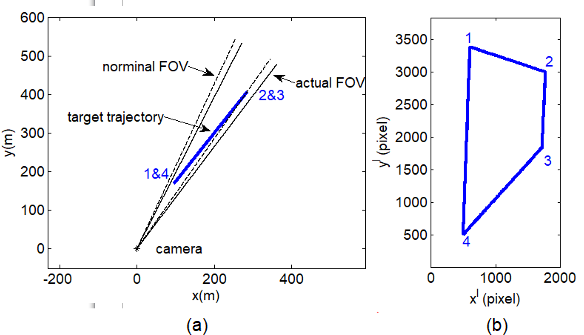}
	\end{center}
	\caption{Test scenario 1. Target moves in constant velocity in a vertical rectangle (1,2,3,4) twice. The higher and lower horizontal edges are at altitudes 284m and 84m, respectively, and near and far vertical edges are at ranges 200m and 500m, respectively. The target speed is 12.5m/s. (a) Top view in the 3D common coordinates. (b) Trajectory in image coordinates.}
	\label{f5}
\end{figure}

\begin{figure}[ht]    
	\begin{center}\leavevmode
		\includegraphics[width=3.4in]{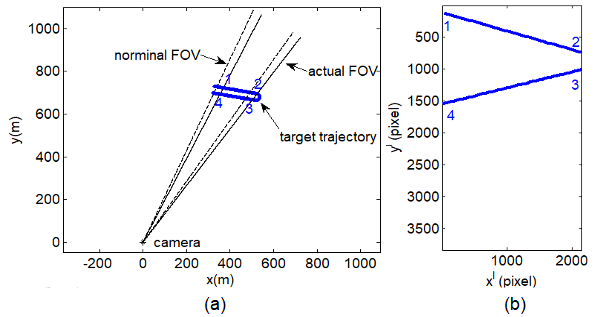}
	\end{center}
	\caption{Test scenario 2. Target moves in constant velocity, makes a 180$^o$ turn, and flies back in constant velocity. The target speed is 12.5m/s. (a) Top view in the 3D common coordinates. (b) Trajectory in image coordinates.}
	\label{f6}
\end{figure}

This section evaluates the performance of the algorithm described in Section~\ref{s4}. We simulate two test scenarios. Scenario~1 shown in Fig.~\ref{f5} has a drone (quadcopter) moving in a vertical rectangular trajectory 1,2,3,4 with two cycles. Points~1 and~4 are at near range 200m with altitudes 284m and 98m, respectively. Points~2 and~3 are at farther range 500m  with altitudes 284m and 98m, respectively. The drone moves with a nearly constant speed of 12.5m/s between the four edges. When reaching a corner, it decelerates to 0m/s, then accelerates to 12.5m/s on the new direction. The total duration is 109.2s with 546 measurements. The design principle of the trajectory for this scenario is to span the entire FOV (with near and far motion, i.e., also in depth). Scenario~2 uses the recommended drone path in~\cite{Yang2022}. This is shown in Fig.~\ref{f6} with the drone moving with speed of 12.8m/s at altitude 100m, and then it makes a 180$^{\rm o}$ turn, and flies back with the same speed and altitude. The total duration is 36s with 126 measurements. The inaccurate GPS trajectories are discretized with a time interval of 0.1s. Camera measurements sampling interval is 0.2s.
The camera to calibrate has a field of view of 10$^{\mathrm o}$ and 17.8$^{\mathrm o}$ horizontal and vertical, respectively. The nominal orientation angles\footnote{The nominal values are the design values. After system installation, the actual values are usually biased w.r.t.\ the nominal values.} are set as $\alpha=30^{\mathrm o}$, $\epsilon=2^{\mathrm o}$ and $\rho=0^{\mathrm o}$. However their actual values (to be estimated) are $\alpha=32^{\mathrm o}$, $\epsilon=4.1^{\mathrm o}$ and $\rho=2.3^{\mathrm o}$. The camera provided the measurements only when the target is in its field of view with measurement error standard deviation $\sigma_{\rm F}$=1pixel. The time and altitude offsets are $\tau=1.35$s and $\hbar=10$m in both scenarios, respectively. We set the time offset precision lower than the GPS discretized precision\footnote{The smallest significant digit for the offset is 0.01s while for the GPS it is 0.1s.} on purpose to observe the algorithm estimation accuracy better. We will study the estimation accuracy, the statistical efficiency through Normalized Estimation Error Squared (NEES) w.r.t.\ the CRLB [2] and the real impact of the results next.

\subsection{Estimation Accuracy}
We conducted 100 Monte Carlo runs for each scenario, and recorded the Root Mean Square Error~(RMSE). The CRLB-based covariance matrix is also computed as a benchmark, and is given by
\begin{equation}
    \bd P=(\bd J'\bd R^{-1}\bd J)^{-1}
\end{equation}
where $\bd J$ is computed by~(\ref{e4_2}), but $\theta$ used in~(\ref{e4_2}) is the true value, namely,
\begin{equation}
    \theta=[32\dg\ 4.1\dg\ 2.3\dg\  10{\rm m}\ 1.35{\rm s}]'
\end{equation}
Note the three angles in $\theta$ should be converted to radians as the unit of measurement in both CRLB and ILS computing, as discussed before. The CRLB standard deviations of the estimated parameters are
\begin{eqnarray}
    \alpha^{\rm{CRLB}}&=&\sqrt{\bd P(1,1)}\\
    \epsilon^{\rm{CRLB}}&=&\sqrt{\bd P(2,2)}\\
    \rho^{\rm{CRLB}}&=&\sqrt{\bd P(3,3)}\\
    \hbar^{\rm{CRLB}}&=&\sqrt{\bd P(4,4)}\\
    \tau^{\rm{CRLB}}&=&\sqrt{\bd P(5,5)}
\end{eqnarray}

Table~\ref{tb_2} gives the RMSE and CRLB for scenarios~1 and~2. It also lists the results of the scenario~2 (under 2*) obtained in~\cite{Yang2022}, where the same drone path was used, but GPS altitude was assumed perfect without bias. It can be seen that the estimate RMSEs of scenario~1 are close to their CRLBs. The algorithm is statistically efficient in this scenario, as shown in the next subsection. However, the results of scenario~2 are significantly less accurate. The CRLBs are significantly larger than those of scenario~1, especially for $\epsilon$ and $\hbar$ with values 32.21mdeg and 475mm, respectively. This indicates the observabilities of $\epsilon$ and $\hbar$ are marginal in this scenario. 

\begin{figure}[ht]
	\begin{center}\leavevmode
		\includegraphics[width=3.0in]{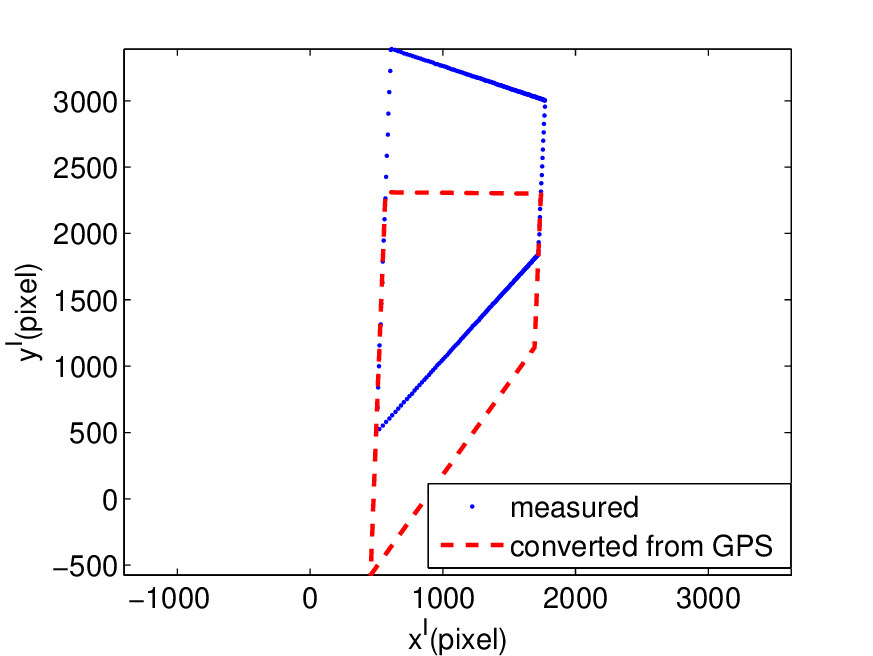}
	\end{center}
	\caption{Measured and GPS converted trajectories of scenario~1, where the actual and nominal yaw, roll and time difference are set to the same values as $\alpha=32\dg$, $\rho=2.3\dg$ and $\tau=1.35$s, respectively. The actual and nominal pitch are $\epsilon=4.1\dg$ and $2\dg$, respectively. The actual and nominal GPS altitude bias are 10m and 0m, respectively.}
	\label{f7}
\end{figure}
\begin{figure}[ht]
	\begin{center}\leavevmode
		\includegraphics[width=3.0in]{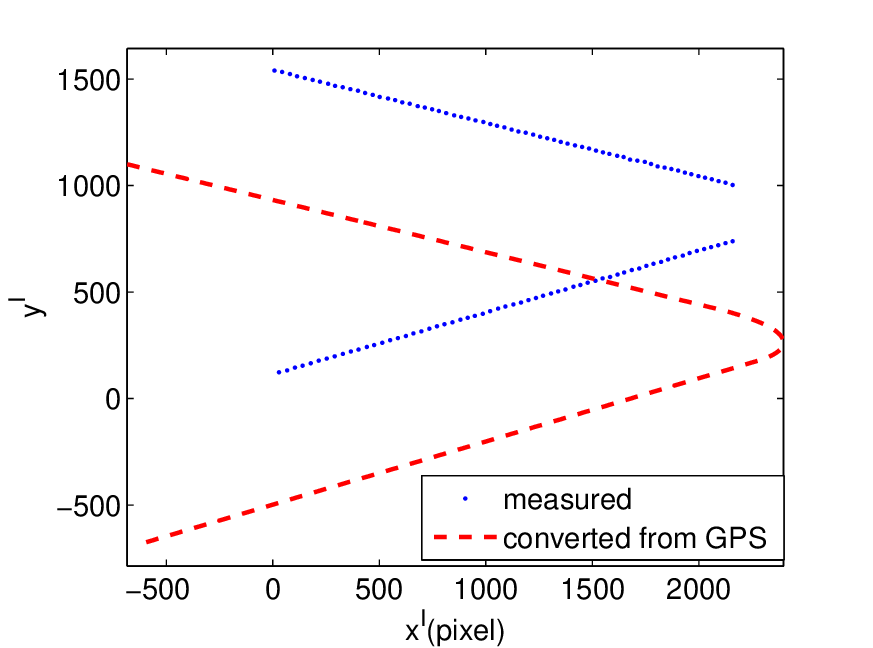}
	\end{center}
	\caption{Measured and GPS converted trajectories of scenario~2, where the actual and nominal yaw, roll and time difference are set to the same values as $\alpha=32\dg$, $\rho=2.3\dg$ and $\tau=1.35$s, respectively. The actual and nominal pitch are $\epsilon=4.1\dg$ and $2\dg$, respectively. The actual and nominal GPS altitude bias are 10m and 0m, respectively.}
	\label{f8}
\end{figure}

We plot the drone trajectories as seen by the camera and the GPS converted positions in the image space for scenarios~1 and~2 in Figs.~\ref{f7} and ~\ref{f8}, respectively. The parameters used for GPS conversion are set the same as the true values, except for the two marginally observable parameters $\epsilon$ and $\hbar$. The true values are $\epsilon=4.1\dg$ and $\hbar=10$m, respectively. The values used in the GPS conversion are $\epsilon=2\dg$ and $\hbar=0$m, respectively. It can be seen that the true and the GPS converted trajectories for scenario~1 (Fig.~\ref{f7}) are quite different. This is mainly because the longer vertical edge is at near range~200m, and the shorter vertical edge is at farther range~500m. One cannot match them without correct values on both~$\epsilon$ and $\hbar$. However, the trajectories for scenario~2 (Fig.~\ref{f8}) are almost parallel. Since the two legs on the drone path are at similar range, one can change either GPS altitude or pitch to match the two trajectories. 
Furthermore, we can also observe that the difference between the RMSE and $\sigma_{\rm CRLB}$ for $\epsilon$ and $\hbar$ are also significantly larger in  scenario~2 than those of scenario~1. The algorithm does not perform well when the problem observability is marginal, as in scenario 2 which does not meet the design principle of Scenario~1.

\begin{table}[!h]
	\caption{CRLB and RMSE from 100 runs}
	\label{tb_2}
	\centering
	\begin{tabular}{c|l|cc}
		 \hline \hline
		 Scenario & Parameter & RMSE & $\sigma_{\rm CRLB}$ \\\hline
         	  & $\alpha$---yaw (mdeg)  &   0.23 & 0.21\\
         	  & $\epsilon$---pitch (mdeg)& 0.87  & 0.82\\
         1 	  & $\rho$---roll (mdeg)    &  2.90 & 2.81\\
          	  & $\hbar$--- (mm)    &  4.69 & 4.35\\
          	  & $\tau$ (ms)      &    0.27  & 0.22\\\hline
          	  & $\alpha$---yaw (mdeg)  &   1.34 & 1.05\\
         2 	  & $\epsilon$---pitch (mdeg)&   36.08 & 32.21\\
          	  & $\rho$---roll (mdeg---yaw)&   14.63 & 14.25\\
          	  & $\hbar$--- (mm)    &  531 & 475\\
          	  & $\tau$ (ms)      &   0.80  & 0.75\\\hline 
          	  & $\alpha$---yaw (mdeg)  &   0.45 & 0.43\\
         2*	  & $\epsilon$---pitch (mdeg)&   0.47 & 0.43\\
          	  & $\rho$---roll (mdeg)    &   8.89 & 8.42\\
          	  & $\tau$ (ms)      &   0.57  & 0.50\\\hline
	\end{tabular}
\end{table}


Comparing scenarios 2 and 2* we can see that including GPS altitude bias (which is generally present) significantly increases the estimation error using the drone path recommended in~\cite{Yang2022}. The path is not practical for camera calibration when GPS altitude bias is taken into consideration. The reason is that the trajectory of scenario~2 has poor observerbility when both GPS altitude and camera pitch are unknown.

Another interesting observation is the estimation accuracy of $\tau$ is smaller than the GPS time discretization of 100ms. The best RMSE reaches 0.27ms in test scenario~1. This shows that the trajectory estimation algorithm described in Sec.~\ref{s4} overcomes the discretization of the GPS trajectory problem effectively.

\subsection{Statistical Efficiency}
\begin{figure}[ht]
	\begin{center}\leavevmode
		\includegraphics[width=3.3in]{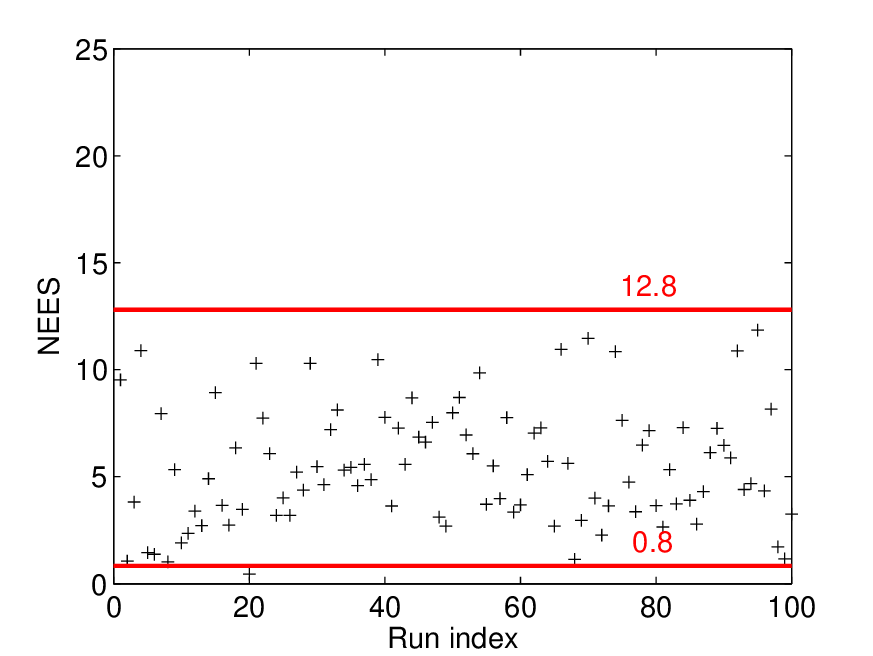}
	\end{center}
	\caption{NEES of 100 runs of the scenario~1.}
	\label{f9}
\end{figure}

\begin{figure}[ht]
	\begin{center}\leavevmode
		\includegraphics[width=3.3in]{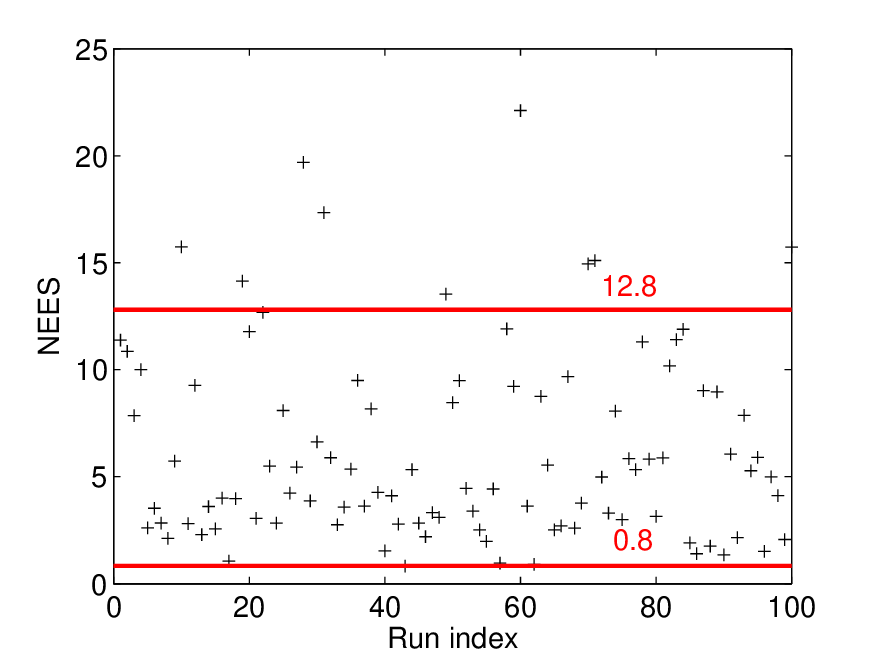}
	\end{center}
	\caption{NEES of 100 runs of the scenario~2.}
	\label{f10}
\end{figure}

The statistical efficiency analysis was conducted using the Normalized Estimation Error Squared (NEES)~\cite{BarShalom11} computed w.r.t.\ the CRLB, namely
\begin{equation}
    \epsilon^i(t_k)=(\theta-\hat{\theta})'\bd P^{-1}(\theta-\hat{\theta})
\end{equation}
where $\hat{\theta}$ and $\theta$ are the parameter estimate and true value, respectively. The NEESs of $N$=100 runs were recorded and the analysis is carried out for each run, as well as using the average. The NEES of the parameter (with dimension 5) is a 5 degrees of freedom chi-square random variable if the errors are Gaussian. Its two-sided $p$=95\% probability region is [0.8, 12.8]. The estimation is  statistically efficient, if 95\% of NEESs are within this region. Figs.~\ref{f9}--\ref{f10} show the NEES for the two test scenarios, and the number of NEES out of the region [0.8, 12.8] for scenario 1 is~0 (vs.\ the expected value of 5), i.e., the algorithm produced statistically efficient estimates --- consistent with equality in the CRLB. However, the number of NEES out of the this interval from 100 runs is  scenario~2 is 9. This shows that the estimation algorithm  for a marginally observable scenario is marginally statistically efficient. This is because the standard deviation of the number of exceedances of the 95\% probability interval is $\sqrt{Np(1-p)}\approx 2$, thus the borderline efficiency.

For the average NEES over 100 runs, the 95\% probability region, based on $\chi^2_{500}/100$, is the interval [4.1\ 5.63]. For scenario 1 the average NEES is 4.69 while for scenario 2 it is 5.86 Thus, the same conclusions can be drawn: for scenario 1 the algorithm is efficient while for scenario 2 it is borderline.

\subsection{Impact of the Residual Biases}

\begin{figure}[ht]
	\begin{center}\leavevmode
		\includegraphics[width=3.6in]{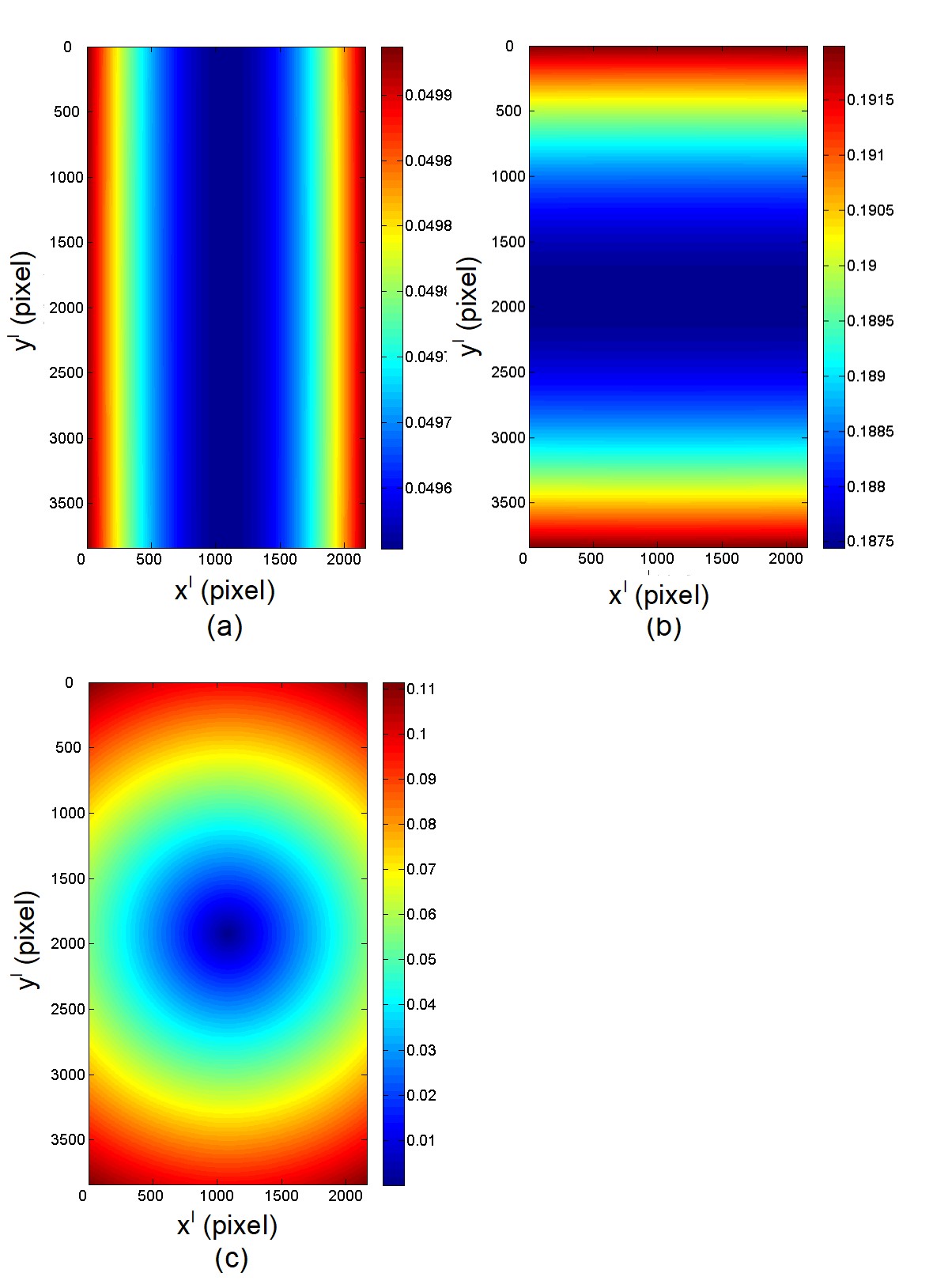}
	\end{center}
	\caption{Biased errors on all 5$\times$5 grids. (a) The biased error caused by calibration error on yaw of 0.23mdeg. (b) The biased error caused by calibration error on pitch of 0.87mdeg. (c) The biased error caused by calibration error on roll of 2.9mdeg.}
	\label{f11}
\end{figure}
\begin{figure}[ht]
	\begin{center}\leavevmode
		\includegraphics[width=3.6in]{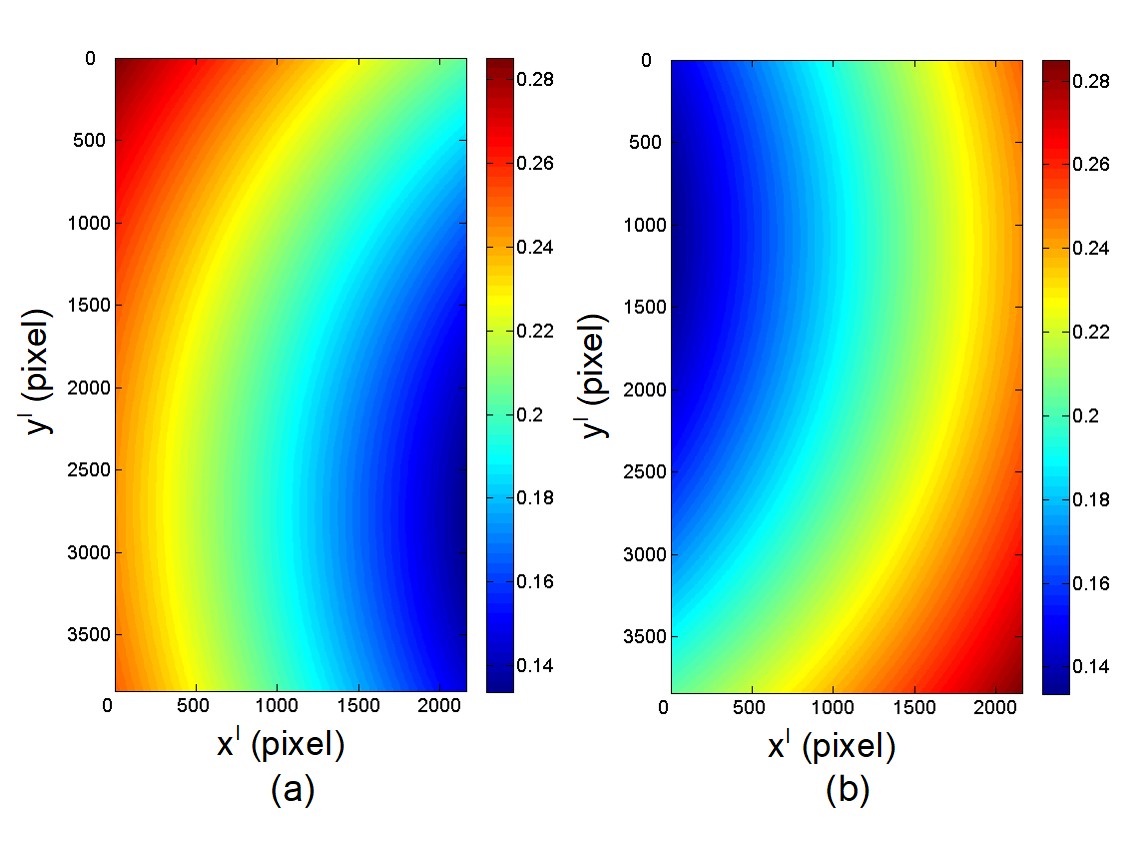}
	\end{center}
	\caption{Biased errors on all 5$\times$5 grids. (a) Increases the yaw, pitch and roll by 0.23mdeg, 0.87mdeg and 2.9mdeg, respectively. (b) Reduces the yaw, pitch and roll by 0.23mdeg, 0.87mdeg and 2.9mdeg, respectively.}
	\label{f12}
\end{figure}
The real impact is further discussed based on the calibration result of scenario~1 which yields a good calibration result. The pixel bias error in the image space caused by the residual calibration errors should be much lower than the measurement error, so that the residual calibration errors are negligible. We compute the pixel bias error based on the calibration RMSE of yaw, pitch, roll and their combination. The residual bias error impact is obtained from the shifted distances (the unit of measure is pixel) for uniformly distributed  5$\times$5 pixel grid elements covering the whole image space\footnote{The errors for each pixel in such a small grid are practically the same, so there is no point in evaluating the impact of the errors in each pixel separately.} (1--2160 in $x\uI$, 1--3840 in $y\uI$) when the residual yaw, pitch and roll errors are introduced. The residual bias error of the $k$th grid is
\begin{equation}
    b_k=|(\breve x\uI_k,\breve y\uI_k)-(x\uI_k,y\uI_k)|
\end{equation}
where $(x\uI_k,y\uI_k)$ is the center of the $k$th grid element in pixel units and 
$(\breve x\uI_k,\breve y\uI_k)$ is the shifted grid center when the residual calibration errors are added to the nominal yaw, pitch and roll; $b_k$ is the distance in pixel units between these two grids. We recorded the residual errors $b_k$ of all the grids and plot them in Fig.~\ref{f11} for three cases. Case~(a) has 0.23mdeg calibration error added to yaw only. Cases~(b) and~(c) have 0.87mdeg error added to pitch and 2.9mdeg error added to roll, respectively. Fig.~\ref{f12} shows the effect of the combination of yaw, pitch and roll errors. Case~(a) increases the yaw, pitch and roll by 0.23mdeg, 0.87mdeg and 2.9mdeg, respectively. Case~(b) reduces the yaw, pitch and roll by 0.23mdeg, 0.87mdeg and 2.9mdeg, respectively. The statistics of the grid biases are summarized in Table~\ref{tb_3}. It shows the bias min., max., mean, standard deviation and Root Mean Square (RMS) value for the five cases in Figs.~\ref{f11} and~\ref{f12}. From these results we observe the following:
\begin{itemize}
    \item The residual bias error is negligible compared to the measurement error. The highest RMSE due to the residual bias is 0.20pixel. The measurement RMSE in one coordinate (either $x\uI$ or $y\uI$) is 1pixel. Assuming they are uncorrelated between the coordinates, the total measurement error standard deviation is 1.41pixel. The highest RMSE due to residual bias is 7.2 times smaller than the measurement RMSE. Thus the calibration using the scenario~1 drone trajectory achieves negligible bias error.
    \item A yaw error creates higher bias on the two vertical edges, and pitch error creates higher bias on the two horizontal edges, as shown in Fig.~\ref{f11}~(a) and (b). The differences between the edges and the center are, however, very small.
    \item A roll error creates higher bias at the four corners, the furthest distance to the center, and the center has zero bias in Fig.~\ref{f11}~(c). Nevertheless, the bias at the corners is negligible.
    \item A combined yaw, pitch and roll error creates the highest bias at one of the corners from Fig.~\ref{f12}. However, the max. 0.29pixels is still negligible compared to the measurement RMSE of 1.41pixel. The max. combined RMSE (measured and bias) is $\sqrt{1.41^2+0.29^2}=1.44$pixel.
\end{itemize}

\begin{table}[!h]
	\caption{Biased error in pixel caused by the calibration error}
	\label{tb_3}
	\centering
	\begin{tabular}{ccc|ccccc}
		 \hline \hline
		  \multicolumn{3}{c|}{Calibration error} & \multicolumn{5}{c}{Bias}\\
		 \multicolumn{3}{c|}{ (mdeg)} & \multicolumn{5}{c}{ (pixel)}\\\hline
		 $\alpha$ & $\epsilon$ & $\rho$ & Min.& Max. & Mean &Stdv. & RMS \\\hline
		 0.23 & 0 & 0 & 0.050 & 0.050 & 0.050 &0.000  & 0.050 \\
		 0 & 0.87 & 0 & 0.187 & 0.192 & 0.189 & 0.001 & 0.189  \\
		 0 & 0 & 2.90 & 0.000 & 0.110 & 0.064 & 0.025 & 0.059  \\
		 0.23 & 0.87 & 2.90 & 0.134 & 0.285 & 0.210 & 0.033 & 0.200\\
		 -0.23 & -0.87 & -2.90 & 0.134 & 0.285 & 0.210 & 0.033 & 0.200\\\hline
	\end{tabular}
\end{table}

\section{Conclusions}
\label{s6}
In this paper, we developed a camera calibration algorithm using drone trajectories recorded by a GPS receiver. However, the recorded GPS data has an unknown altitude bias and an unknown time offset between the GPS and camera systems. The GPS trajectories are discretized with time interval 0.1s. The paper developed a special ML/ILS algorithm dealing with discretized GPS trajectories to estimate camera orientation angles (yaw, pitch and roll), GPS altitude bias and time offset simultaneously. The simulation tests were conducted and an appropriate drone trajectory is recommended whose estimation results met the CRLB and NEES requirements. The time offset estimation error was much smaller than the discretization of the GPS reference trajectory (0.27ms vs.\ 100ms). The recommended drone trajectory is suitable for practical use. Its residual calibration bias RMSE was 14\% of the measurement error standard deviation, which is negligible.

In our real camera setup and calibration experiments, we realized that more work needs to be done along this research. First, the camera focal length cannot be fixed beforehand accurately. It needs to be adjusted during setup based on the real situation. Due to lack of accurate equipment to measure a camera focal length, it should be an additional camera parameter included in the estimation. Second, the GPS equipment usually has quantization error in latitude and longitude. This error cannot be ignored when a target is in a near range (with 10 pixel quantization). The ILS algorithm proposed in this paper need to be further developed to handle these types of errors.

\appendices
\section{The Importance of Being Earnest about Radians}
\label{a1}
When trigonometric functions are expressed as Taylor expansion, one has to use radians as the unit of measure. This can be illustrated using the following simple example, using the first order Taylor expansion to compute $\sin(30.01\dg)$. The answer should be~0.50015. If we use degrees as the unit of measure we will have wrong result as
\begin{align}
    \sin(30.01\dg)= & \sin(30\dg+0.01\dg)\nonumber\\
    \approx & \sin(30\dg)+0.01\dg\times[\sin(30\dg)]'\nonumber\\
    \approx & \sin(30\dg)+0.01\dg \times \cos(30\dg)\nonumber\\
    \approx & 0.5+0.01\times 0.866\nonumber\\
    \approx & 0.50866\label{e4_23}
\end{align}
If we use radians, the correct result is
\begin{align}
    \sin\left(\dfrac{30.01\times \pi}{180}\right)\approx & \sin\left(\dfrac{30\times \pi}{180}\right)+\dfrac{0.01\times \pi}{180} \cos\left(\dfrac{30\times \pi}{180}\right)\nonumber\\
    \approx & 0.5+0.00175\times 0.866\nonumber\\
    \approx & 0.50015\label{e4_24}
\end{align}
Although $\sin(\cdot)$ and $\cos(\cdot)$ should give the same values whether the units are degrees or radians, the small difference $0.01\dg$ in front of $\cos(\cdot)$ in (\ref{e4_23}) leads to wrong result in~(\ref{e4_24}). Thus angles must be converted to radians when using series expansions.

\section{Derivatives for (\ref{e4_4})}
\label{a2}
The iteration index $j$ is omitted for simplicity. The gradients needed are
\begin{equation}
    [\nabla_{\bd \theta} \bd h_k(\cdot)']'=\dfrac{\partial\bd x_k\uI}{\partial\bd x_k\uC}\dfrac{\partial\bd x_k\uC}{\partial\theta}\ \ \ \ \ \ k=1\ldots n\label{e4_5}
\end{equation}

\begin{eqnarray}
    \dfrac{\partial\bd x_k\uI}{\partial\bd x_k\uC}&=&\left[\begin{array}{ccc}
    \dfrac{f}{z_k\uC} & 0 & -\dfrac{f x_k\uC}{(z_k\uC)^2}\\
    0     & \dfrac{f}{z_k\uC}  & -\dfrac{f y_k\uC}{(z_k\uC)^2}
    \end{array}\right]\label{e4_6}\\
    \dfrac{\partial\bd x_k\uC}{\partial\theta}&=&\left[\begin{array}{ccccc}
        \dfrac{\partial x_k\uC}{\partial \alpha} & \dfrac{\partial x_k\uC}{\partial \epsilon} & \dfrac{\partial x_k\uC }{\partial \rho} & \dfrac{\partial x_k\uC }{\partial \hbar} & \dfrac{\partial x_k\uC}{\partial \tau}\\
        \dfrac{\partial y_k\uC}{\partial \alpha} & \dfrac{\partial y_k\uC}{\partial \epsilon} & \dfrac{\partial y_k\uC}{\partial \rho} & \dfrac{\partial y_k\uC }{\partial \hbar} & \dfrac{\partial y_k\uC}{\partial \tau}\\
        \dfrac{\partial z_k\uC}{\partial \alpha} & \dfrac{\partial z_k\uC}{\partial \epsilon} & \dfrac{\partial z_k\uC}{\partial \rho} & \dfrac{\partial z_k\uC }{\partial \hbar} & \dfrac{\partial z_k\uC}{\partial \tau}
    \end{array}\right]\label{e4_7}
\end{eqnarray}
and
\begin{eqnarray}
\dfrac{\partial x_k\uC}{\partial \alpha}&=&\Delta x_k(\cosa\sine\sinp-\sina\cosp)-\Delta y_k(\sina\sine\sinp+\cosa\cosp)\nonumber\\\label{e4_8}\\
\dfrac{\partial x_k\uC}{\partial \epsilon}&=&\Delta x_k\sina\cose\sinp+\Delta y_k\cosa\cose\sinp+\Delta z_k\sine\cosp\label{e4_9}\\
\dfrac{\partial x_k\uC}{\partial \rho}&=&\Delta x_k(\sina\sine\cosp-\cosa\sinp)\nonumber\\
    &&+\Delta y_k(\cosa\sine\cosp+\sina\sinp)-\Delta z_k\cose \cosp\label{e4_10}\\
\dfrac{\partial x_k\uC}{\partial \hbar}&=&\cose\sinp\\    
\dfrac{\partial x_k\uC}{\partial \tau}&=&\hat{\dot{x}}(t_k+\tau)(\cosa\cosp+\sina\sine\sinp)\nonumber\\
    &&+\hat{\dot{y}}(t_k+\tau)(\cosa\sine\sinp-\sina\cosp)-\hat{\dot{z}}(t_k+\tau)\cose\sinp\nonumber\\
    \label{e4_11}
\end{eqnarray}
\begin{eqnarray}
\dfrac{\partial y_k\uC}{\partial \alpha}&=&\Delta x_k(\cosa\sine\cosp+\sina\sinp)+\Delta y_k(\cosa\sinp-\sina\sine\cosp)\nonumber\\\label{e4_12}\\
\dfrac{\partial y_k\uC}{\partial \epsilon}&=&\Delta x_k\sina\cose\cosp+\Delta y_k\cosa\cose\cosp+\Delta z_k\sine\cosp\label{e4_13}\\
\dfrac{\partial y_k\uC}{\partial \rho}&=&-\Delta x_k(\sina\sine\sinp+\cosa\cosp)\nonumber\\
    &&+\Delta y_k(\sina\cosp-\cosa\sine\sinp)+\Delta z_k\cose\sinp\label{e4_14}\\
\dfrac{\partial y_k\uC}{\partial \hbar}&=&\cose\cosp\\    
\dfrac{\partial y_k\uC}{\partial \tau}&=&\hat{\dot{x}}(t_k+\tau)(\sina\sine\cosp-\cosa\sinp)\nonumber\\
    &&+\hat{\dot{y}}(t_k+\tau)(\sina\sinp+\cosa\sine\cosp)-\hat{\dot{z}}(t_k+\tau)\cose\cosp\nonumber\\
    \label{e4_15}
\end{eqnarray}    
\begin{eqnarray}
    \dfrac{\partial z_k\uC}{\partial \alpha}&=&\Delta x_k\cosa\cose-\Delta y_k\sina\cose\label{e4_16}\\
    \dfrac{\partial z_k\uC}{\partial \epsilon}&=&-\Delta x_k\sina\sine-\Delta y_k\cosa\sine+\Delta z_k\cose\label{e4_17}\\
    \dfrac{\partial z_k\uC}{\partial \rho}&=&0\label{e4_18}\\
    \dfrac{\partial x_k\uC}{\partial \hbar}&=&-\sine\\        
    \dfrac{\partial z_k\uC}{\partial \tau}&=&\hat{\dot{x}}(t_k+\tau)\sina\cose+\hat{\dot{y}}(t_k+\tau)\cosa\cose+\hat{\dot{z}}(t_k+\tau)\sine\nonumber\\\label{e4_19}    
\end{eqnarray}
where
\begin{eqnarray} 
    \Delta x_k&=&\hat x(t_k+\tau)-x\us\label{e4_20}\\
    \Delta y_k&=&\hat y(t_k+\tau)-y\us\label{e4_21}\\
    \Delta z_k&=&\hat z(t_k+\tau)-\hbar-z\us\label{e4_22}
\end{eqnarray}

The point [$\hat x(t_k+\tau)$,~$\hat y(t_k+\tau)$,~$\hat z(t_k+\tau)$] in~(\ref{e4_20})--(\ref{e4_22}) on the drone trajectory and its velocity [$\hat{\dot{x}}(t_k+\tau)$, $\hat{\dot{y}}(t_k+\tau)$, $\hat{\dot{z}}(t_k+\tau)$] in~(\ref{e4_11}),~(\ref{e4_15}) and~(\ref{e4_19}) have been estimated in~\ref{s41}.

The unit of measure for the three angles $\alpha$, $\epsilon$ and $\rho$ has to be radians --- see Appendix~\ref{a1}.

\begin{IEEEbiography}[{\includegraphics[width=1in]{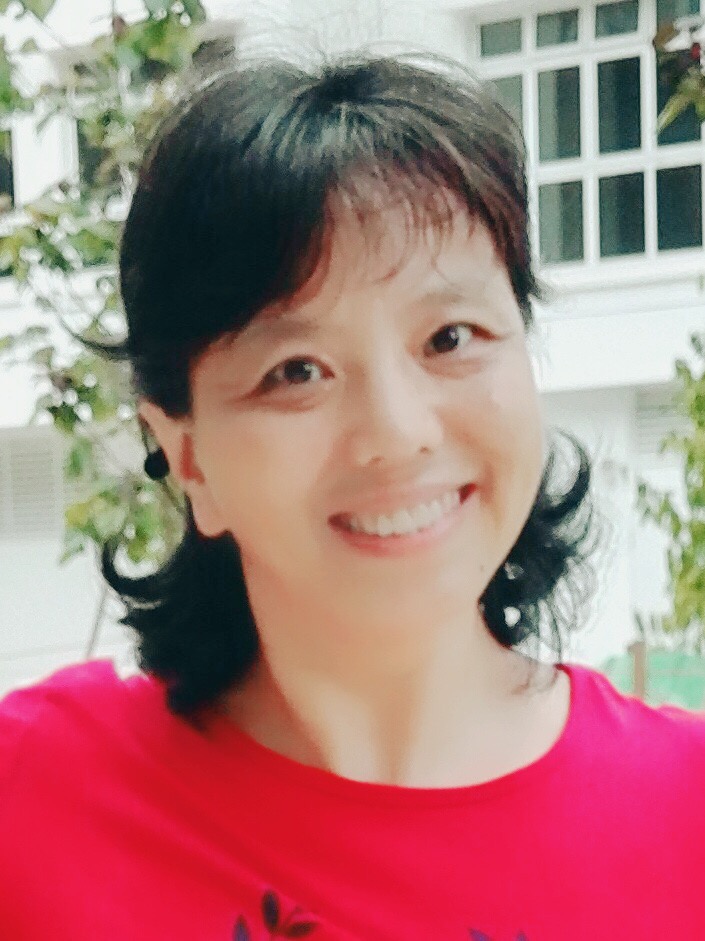}}]{Rong Yang}
	received her B.E. degree in information and control from Xi'an Jiao Tong University, China in 1986, M.Sc. degree in electrical engineering from National University of Singapore in 2000, and Ph.D. degree in electrical engineering from Nanyang Technological University, Singapore in 2012. She is currently a Principal Member of Technical Staff at DSO National Laboratories, Singapore. Her research interests include passive tracking, low observable target tracking, GMTI tracking, hybrid dynamic estimation and data fusion. She was Publicity and Publication Chair of FUSION 2012 and received the FUSION 2014 Best Paper Award (First runner up).
\end{IEEEbiography}

\begin{IEEEbiography}[{\includegraphics[width=1in]{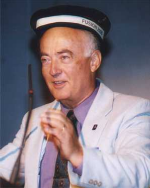}}]{Yaakov Bar-Shalom} (F'84)  received the B.S. and M.S. degrees from the Technion in 1963 and 1967 and the Ph.D. degree from Princeton University in 1970, all in EE. Currently he is Board of Trustees Distinguished Professor in the ECE Dept. and Marianne E. Klewin Professor at the University of Connecticut. His current research interests are in estimation theory, target tracking and data fusion. He has published over 650 papers and book chapters. He coauthored/edited 8 books, including Tracking and Data Fusion (YBS Publishing, 2011). He has been elected Fellow of IEEE for "contributions to the theory of stochastic systems and of multitarget tracking". He served as Associate Editor of the IEEE Transactions on Automatic Control and Automatica. He was General Chairman of the 1985 ACC, General Chairman of FUSION 2000, President of ISIF in 2000 and 2002 and Vice President for Publications during 2004-13. Since 1995 he is a Distinguished Lecturer of the IEEE AESS. He is corecipient of the M. Barry Carlton Award for the best paper in the IEEE TAE Systems in 1995 and 2000. In 2002 he received the J. Mignona Data Fusion Award from the DoD JDL Data Fusion Group. He is a member of the Connecticut Academy of Science and Engineering. In 2008 he was awarded the IEEE Dennis J. Picard Medal for Radar Technologies and Applications, and in 2012 the Connecticut Medal of Technology. He has been listed by  academic.research.microsoft (top authors in engineering) as \#1 among the researchers in Aerospace Engineering based on the citations of his work. He is the recipient of the 2015 ISIF Award for a Lifetime of Excellence in Information Fusion. This award has been renamed in 2016 as the Yaakov Bar-Shalom Award for a Lifetime of Excellence in Information Fusion. He has the following Wikipedia page: https://en.wikipedia.org/wiki/Yaakov Bar-Shalom.
\end{IEEEbiography}

\begin{IEEEbiography}[{\includegraphics[width=1in]{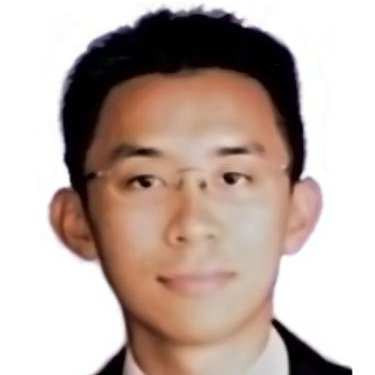}}]{Huang Hong'An Jack}
was born in Singapore in 1983. He received the B.E. from National University of Singapore (NUS) in 2008. He is currently a Senior Member of Technical Staff at DSO National Laboratories. His research interests in target tracking including GMTI tracking, passive tracking and image tracking. He received the FUSION 2014 Best Paper Award (First runner up).
\end{IEEEbiography}

\end{document}